\documentclass[10pt, final]{IEEEtran}

\usepackage{url}
\usepackage[cmex10]{amsmath}
\usepackage{amsfonts}
 \usepackage{theorem}
\usepackage[dvips]{epsfig}
\usepackage{graphicx}

 \usepackage{verbatim}
 \usepackage{amssymb}
\usepackage{amsbsy}
\usepackage{url}

\theoremstyle{plain}
\newtheorem{Theorem}{Theorem}

\newtheorem{Proposition}{Proposition}
\newtheorem{Corollary}{Corollary}

\newtheorem{Problem}{Problem}

{\theorembodyfont{\rmfamily} \newtheorem{Remark}{Remark}}
{\theorembodyfont{\rmfamily} }
{\theorembodyfont{\rmfamily} \newtheorem{Example}{Example}}
{\theorembodyfont{\rmfamily} }
{\theorembodyfont{\rmfamily} }

\newcommand {\R}{\mathbb R}

\newcommand{\be}{\begin{equation}}
\newcommand{\ee}{\end{equation}}

\begin{document}
%
%
\title{Maximizing Protein Translation Rate in the
 Ribosome Flow Model: the Homogeneous Case\thanks{This research   is partially supported by research grants
from  the  ISF and from the Ela Kodesz  Institute for Medical Engineering and Physical Sciences.}}
\author{Yoram Zarai, Michael Margaliot and Tamir Tuller \IEEEcompsocitemizethanks{\IEEEcompsocthanksitem Y.
Zarai is with the School of Electrical Engineering, Tel-Aviv
University, Tel-Aviv 69978, Israel. \protect\\
E-mail: yoramzar@mail.tau.ac.il
\IEEEcompsocthanksitem
M. Margaliot is with the School of Electrical Engineering and the Sagol School of Neuroscience, Tel-Aviv
University, Tel-Aviv 69978, Israel. \protect\\
E-mail: michaelm@eng.tau.ac.il \IEEEcompsocthanksitem T. Tuller is
with the Department of Biomedical Engineering and the Sagol School of Neuroscience,  Tel-Aviv
University, Tel-Aviv 69978, Israel. \protect\\
E-mail: tamirtul@post.tau.ac.il }
\thanks{}}

\maketitle


\begin{abstract}
Gene translation is the process in which intracellular macro-molecules, called
ribosomes,   decode
 genetic information in the mRNA chain into the corresponding proteins.
 Gene translation includes several steps. During the
  elongation step, ribosomes   move
  along the mRNA   in a sequential manner and
  link amino-acids together in the corresponding order to produce the proteins.

The homogeneous ribosome flow model~(HRFM) is a   deterministic computational
 model for translation-elongation under the assumption
  of   constant elongation rates along the mRNA chain.
  The HRFM is  described by a set of~$n$ first-order nonlinear ordinary differential equations, where~$n$ represents the number of  sites along the mRNA chain. The HRFM   also  includes   two positive parameters:
  ribosomal initiation rate and the (constant) elongation rate.

In this paper, we show that the steady-state translation rate in the HRFM
 is a \emph{concave} function of its parameters.
 This means that the problem of determining the
 parameter values that maximize the translation rate
 is relatively simple. Our results may
 contribute to a better understanding of the mechanisms and evolution
of  translation-elongation.
We demonstrate this by using   the theoretical results
to estimate  the initiation rate in {\em M. musculus} embryonic stem cell.
The underlying assumption is that evolution optimized the translation mechanism.

 For the infinite-dimensional HRFM, we derive a \emph{closed-form} solution
 to the problem of determining the initiation and transition rates that
 maximize the protein translation  rate. We show that these expressions
 provide good approximations for the optimal values in the $n$-dimensional
 HRFM already for relatively small values of~$n$.
These results may have
  applications for synthetic biology where an important problem is to re-engineer  genomic systems in order to
   maximize the protein production rate.
\end{abstract}


\section{Introduction}

Proteins are micro-molecules involved in all intracellular activities. DNA regions, called genes, encode proteins as  ordered lists of amino acids.
 During the process of gene expression these regions are first transcribed into mRNA molecules.
 In the next step, called \emph{gene translation},    the information encoded in the mRNA
  is  translated into proteins by molecular machines called \emph{ribosomes} that move along the mRNA sequence~\cite{Alberts2002}.
 During the translation process, each triplet of consecutive nucleotides, called a  codon, is decoded by a ribosome into a suitable amino-acid.

Gene translation   is a fundamental cellular process
and its study has important implications to numerous  scientific
disciplines ranging from human health to evolutionary biology.
Computational models of translation are becoming increasingly
more important due to the need
  to integrate, analyze, and understand
the rapidly accumulating
 biological findings related to translation~\cite{Zhang1994,Dana2011,Heinrich1980,MacDonald1968,TullerGB2011,Tuller2007,Chu2012}.

Computational models of translation are
also of importance  in synthetic biology.
Indeed, a major challenge in this field is
 to re-engineer  genomic systems to
  produce a desired protein translation rate.
Computational models of translation are   crucial in achieving this goal, as they allow
to     simulate and    analyze the effect of various manipulations of the genomic mechanism
on the translation rate.

A standard mathematical model for translation-elongation is
the \emph{Totally Asymmetric Simple Exclusion Process}~(TASEP)~\cite{Shaw2003,TASEP_tutorial_2011,Lakatos2003,nems2011}.
 TASEP is a stochastic model for   particles moving along a track. A
  chain of sites models the tracks. Each site can be either empty or occupied by a particle.
 The term \emph{simple exclusion} refers to
 the fact that particles hop randomly from one site to the next, but
   only if the target site  is not already occupied. In this way, TASEP encapsulates  the \emph{interaction}
    between the particles.
      The term \emph{totally asymmetric} is used to indicate unidirectional motion along the lattice. Despite its rather simple description,  it seems that rigorous analysis of   TASEP is non-trivial.
See~\cite{TASEP_book} for a detailed exposition of these issues.

In 2011, Reuveni et al.~\cite{reuveni}   considered a  \emph{deterministic} mathematical model
for translation-elongation called the \emph{ribosome flow model}~(RFM).
This model may be derived as a mean-field approximation of      TASEP
(see, e.g.~\cite[p.~R345]{solvers_guide}).

Recent biological
studies have shown  that in some cases the
elongation rates along the mRNA
 are approximately constant~\cite{Ingolia2011,Qian2012}. Under the assumption
of constant elongation rates, the RFM becomes the
\emph{homogeneous ribosome flow model} (HRFM)~\cite{HRFM_steady_state}.
This model includes two positive parameters: the initiation rate~$\lambda$ and the
constant elongation rate~$\lambda_c$.
In this paper, we
show  that  the steady-state  translation rate in the HRFM
 is a  \emph{concave} function of these parameters.
 This implies that the problem of optimizing the translation rate under
 a  simple constraint on the rates is a \emph{convex optimization problem}.
 Thus, this problem admits a unique solution, and this solution can be easily
 found (numerically) using simple and efficient algorithms.
 We also derive an \emph{explicit} expression for the optimal solution
  for the particular case of the infinite-dimensional~HRFM, that is, when~$n\to\infty$.

 These results may
 have important applications in the context of synthetic biology. Indeed, a fundamental
  problem
  in this field is to re-engineer a genetic system  by manipulating  the transcript sequence,
   and possibly other intra-cellular variables, in order to maximize the  translation rate.
  Also, it is reasonable to expect that
  in most organisms evolutionary forces act to optimize translation costs.
 For example, in micro-organisms the growth rate is globally strongly dependent on the translation rate/efficiency (see, for example, \cite{Kudla2009,Tuller2010,dos-Reis2004,dos-Reis2009}). In
   addition, it has been shown that in all organisms  highly expressed genes undergo selection for sequence features that improve their translation rate efficiency (see, for example,  \cite{Kudla2009,Tuller2010,Lithwick2003}).
     The mathematical results described here may be applied to study these issues in a rigorous manner.

 Concavity of  the  translation  rate with respect
  to various variables   can also be examined   experimentally.
  A recent paper~\cite{Firczuk2013}
  studied the effect of the intracellular translation factor abundance on
protein synthesis.
Experiments based on a tet07 construct were used  to manipulate
the   production of the encoded translation factor to a sub-wild-type level~\cite{Firczuk2013},
  and
  measure
   the translation rate, or protein levels,  for each level of the translation factor(s). Their
  results suggest
 that the mapping from levels of translation factors to
  translation rate is indeed concave (see Fig.~$1$ in \cite{Firczuk2013}).
Our  results     thus provide the first mathematical support of the observed concavity in the experiments of~\cite{Firczuk2013}.

The remainder of this paper is organized as follows. Section~\ref{sec:rev}
briefly reviews the RFM and HRFM. Section~\ref{sec:main}   presents the main results.
Section~\ref{sec:bio} describes an application of the theoretical results
for estimating the initiation rate in {\em M. musculus} embryonic stem cell.
The underlying assumption is that evolution optimized the translation mechanism.
The final section summarizes and describes several
possible directions for further research. In order to  streamline the presentation, all the proofs are placed in the Appendix.

\section{Preliminaries}\label{sec:rev}

In the RFM, mRNA molecules are coarse-grained into~$n$ consecutive sites. The RFM
 is given by $n$ first-order nonlinear ordinary differential equations:
\begin{align}\label{eq:rfm}
                    \dot{x}_1&=\lambda (1-x_1) -\lambda_1 x_1(1-x_2), \nonumber \\
                    \dot{x}_2&=\lambda_{1} x_{1} (1-x_{2}) -\lambda_{2} x_{2} (1-x_3) , \nonumber \\
                    \dot{x}_3&=\lambda_{2} x_{ 2} (1-x_{3}) -\lambda_{3} x_{3} (1-x_4) , \nonumber \\
                             &\vdots \nonumber \\
                    \dot{x}_{n-1}&=\lambda_{n-2} x_{n-2} (1-x_{n-1}) -\lambda_{n-1} x_{n-1} (1-x_n), \nonumber \\
                    \dot{x}_n&=\lambda_{n-1}x_{n-1} (1-x_n) -\lambda_n x_n.
\end{align}
Here,~$x_i:\R_+ \to [0,1]$ is the occupancy level at site~$i$ at time~$t$,
normalized so that~$x_i(t)=0$ [$x_i(t)=1$]
implies that site~$i$ is completely empty [completely full] at time~$t$.
The parameter~$\lambda > 0$ is
the initiation rate into the chain,
and~$\lambda_i > 0, i\in\{1,..,n\},$ is a parameter that controls the transition rate from site~$i$ to site~$i+1$. In particular, $\lambda_n$   controls the output rate at the end of the chain.

\begin{figure*}[t]
 \centering
 \scalebox{0.75}{\input{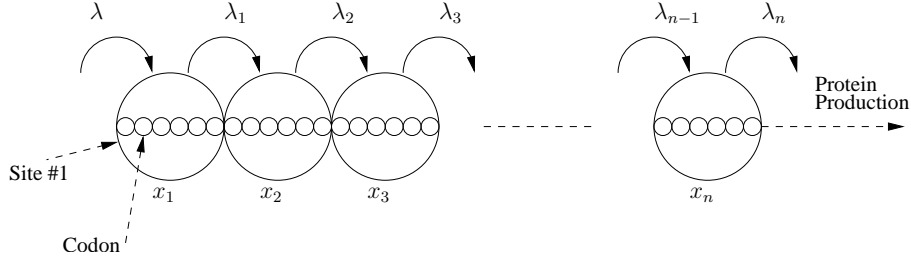}}
\caption{The RFM:
Codons are grouped into sites; $x_i(t)$ is the occupancy level at site~$i$ at time~$t$;
the~$\lambda_i$s control the transition rates between consecutive sites;
the protein production rate at time~$t$ is~$\lambda_n x_n(t)$.}\label{fig:rfmm}
\end{figure*}

The rate of ribosome flow into
the system is $\lambda (1-x_{1}(t))$.
The rate of ribosome flow exiting
the last site, i.e., the \emph{protein translation rate}, is~$\lambda_{n}  x_{n}(t)$.
The    rate of ribosome
flow from site~$i$ to site~$i+1$ is~$\lambda_{i} x_{i}(t)
(1 - x_{i+1}(t) )$  (see Fig.~\ref{fig:rfmm}).
Note that this rate increases with~$x_i(t)$ (i.e., when   site~$i$ is fuller)
and decreases with~$x_{i+1}(t)$ (i.e., when  the consecutive site is becoming fuller).
In this way, the RFM, just like the TASEP,
 takes into account the \emph{interaction} between the ribosomes in consecutive
  sites.

Let~$x(t,a)$ denote the solution of the RFM at time~$t$ for the initial
condition~$x(0)=a$. Since the  state-variables correspond to normalized occupation levels,
  we always consider initial conditions~$a$ in the closed $n$-dimensional
  unit cube:
$
           C^n:=\{x \in \R^n: x_i \in [0,1] , i=1,\dots,n\}.
$
It is straightforward to verify that this implies that~$x(t,a) \in C^n$ for all~$t\geq0$ (see~\cite{RFM_stability}).

Let~$Int(C^n)$ denote the interior of~$C^n$.
It was shown in~\cite{RFM_stability} that the RFM is a
\emph{monotone dynamical system}~\cite{hlsmith}
and that this implies that~\eqref{eq:rfm}
admits a unique equilibrium point~$e \in Int(C^n)$. Furthermore,
$
 \lim_{t\to \infty}x(t,a)=e
$
for all~$a \in C^n$. This means that
all trajectories converge to the steady-state~$e$.
From a biological viewpoint, this means that the ribosome distribution profile along the chain
converges to a steady-state profile that does not depend on the initial profile, but only on
the parameter values.

We note in passing
 that monotone dynamical systems have recently found many
applications in systems biology, see e.g.~\cite{angeli_sontag_pnas_2004,mono_chem_2007,sontag_near_2007}
 and the references therein.

Let
\be \label{eq:defr}
R:=\lambda_n  {e}_n
\ee
denote the \emph{steady-state translation rate}. An important problem is to understand
the dependence of~$e$ and, in particular,~$R$ on the RFM parameters.
For~$x=e$ the left-hand side of all the equations
in~\eqref{eq:rfm} is zero, so
\begin{align} \label{eq:ep}
                      \lambda (1- {e}_1) & = \lambda_1 {e}_1(1- {e}_2)\nonumber \\&
                      = \lambda_2  {e}_2(1- {e}_3)   \nonumber \\ & \vdots \nonumber \\
                    &= \lambda_{n-1} {e}_{n-1} (1- {e}_n) \nonumber \\& =\lambda_n  {e}_n.
\end{align}
This yields
\begin{align}\label{eq:rall}
R=\lambda_i e_i(1-e_{i+1}), \quad i\in\{1,\dots,n\},
\end{align}
where we define~$e_{n+1}:=0$. Also,
\begin{align}\label{eq:list}
                             {e}_n & = R/\lambda_n, \nonumber  \\
                             {e}_{n-1} & = R / (\lambda_{n-1} (1- {e}_n) ),  \nonumber\\
                            & \vdots \nonumber \\
                             {e}_{2} & = R / (\lambda_{2} (1- {e}_3) ), \nonumber\\
                             {e}_{1} & = R / (\lambda_{1} (1- {e}_2) ),
\end{align}
and \be \label{eq:also}
                             {e}_1= 1-\frac{R}{ \lambda} .
\ee
Combining~\eqref{eq:list} and~\eqref{eq:also} provides a
 finite continued fraction~\cite{waad} expression for~$R$:

\begin{align} \label{eq:cf}
                1-R/\lambda&= \cfrac{R/ \lambda_1 }
                                  {  1-\cfrac{R / \lambda_2}
                                  {1-\cfrac{R / \lambda_3 }{\hphantom{aaaaaaa} \ddots
                             \genfrac{}{}{0pt}{0}{}
                             {1-\cfrac{R/\lambda_{n-1}}{1-R/ \lambda_n.}} }}}
\end{align}
Note that \eqref{eq:cf}  has multiple solutions for~$R$ (and thus also
multiple solutions for~$e_n=R/\lambda_n)$, however, we are interested only in the unique feasible solution, i.e. the solution corresponding
 to~$e \in Int(C^n)$.



Recent
biological findings suggest that in some cases
the transition rate along the mRNA chain is approximately constant \cite{Ingolia2011}; this may be also the case for gene transcription~\cite{Edri2013}.
To model this case, Ref.~\cite{HRFM_steady_state} has considered the RFM in the special case where
\be\label{eq:defhrfm}
            \lambda_1=\lambda_2=\dots=\lambda_n := \lambda_c,
\ee
that is, the transition rates~$\lambda_i$ are all equal,
and~$\lambda_c$ denotes their common value. Since this \emph{Homogeneous Ribosome Flow Model}~(HRFM)
 includes only  two parameters,   $\lambda$ and~$\lambda_c$,  the analysis is simplified.
In particular,~\eqref{eq:cf} becomes
\begin{align} \label{eq:cfper}
                R/\lambda&=  1- \cfrac{R/ \lambda_c }
                                  {  1-\cfrac{R / \lambda_c}
                                  {1-\cfrac{R / \lambda_c }{\hphantom{aaaaaaa} \ddots
                             \genfrac{}{}{0pt}{0}{}
                             {1-\cfrac{R/\lambda_c}{1-R/ \lambda_c  }} }}}
\end{align}
where~$\lambda_c$ appears a total of~$n$ times.
Note that the right-hand side here
is  a \emph{1-periodic} continued fraction~\cite{waad}.
Eq.~\eqref{eq:cfper} yields a polynomial equation  of degree~$\lceil (n+1)/2 \rceil$
in~$R$.
For example, for~$n=2$ Eq.~\eqref{eq:cfper} becomes
\begin{align} \label{eq:rn2}
                R^2-(2\lambda+\lambda_c)R+\lambda\lambda_c=0.
\end{align}

Several recent papers analyzed  the RFM/HRFM.
In~\cite{RFM_entrain} it has been shown that
the state-variables (and thus the translation rate) in the RFM
\emph{entrain} to periodically time-varying initiation and/or transition rates.
This provides a  computational framework for studying
entrainment to a periodic excitation, e.g., the $24$ hours solar day or the cell-cycle,
at the genetic level.
Ref.~\cite{RFM_feedback} has considered the
 RFM with positive linear feedback as a model for ribosome recycling.
  It has been
  shown that the closed-loop system admits a unique globally asymptotically stable equilibrium point.
Ref.~\cite{zarai_infi} has considered the HRFM in the case of an infinitely-long chain,
(i.e. when $n\to\infty$) and derived a simple expression for $e_\infty:=\lim_{n\to \infty}e_n$,
as well as  bounds for $|e_\infty-e_n|$ for all~$n\geq 2$.

Summarizing, the   RFM  is a  deterministic  model for translation-elongation,
and perhaps also other
 stages of gene expression~\cite{Zur2012,Edri2013},
 that
is highly amenable to analysis.

In the HRFM, the steady-state
translation rate~$R$ is a function of the positive
parameters~$\lambda,\lambda_c$, i.e.~$R=R(\lambda,\lambda_c)$.
In this paper, we study the dependence of~$R$ on these parameters.

\section{Main results}\label{sec:main}
Our first result
shows that~$R$ is a concave function.
\subsection{Concavity}
\begin{Theorem}\label{thm:concave}
Consider the HRFM with  dimension~$n\geq 2$. The steady-state
translation rate~$R=R(\lambda,\lambda_c)$
is a concave function on~$\R^2_+$.
\end{Theorem}

The next example demonstrates Theorem~\ref{thm:concave} for
the particular case~$n=2$.
\begin{Example}\label{exa:con2}
Consider the HRFM with~$n=2$.
In this case, the feasible solution of~\eqref{eq:list} and \eqref{eq:also}
  (i.e., the solution satisfying~$e_2 \in (0,1)$ for all~$\lambda,\lambda_c>0$) is
\be\label{eq:e2here}
                e_2(\lambda,\lambda_c)=
                 ( 2\lambda+\lambda_c-\sqrt{ 4\lambda^2+\lambda_c^2 }      ) /(2\lambda_c),
\ee
so
\be\label{eq:rfunc}
                R(\lambda,\lambda_c) =
                 ( 2\lambda+\lambda_c-\sqrt{ 4\lambda^2+\lambda_c^2 }      ) /2.
\ee
 It is useful to demonstrate    Theorem~\ref{thm:concave} for this special case.
By~\eqref{eq:rfunc},
$
                \frac{\partial}{\partial \lambda} R=
                 1- \frac{ 2\lambda }{\sqrt{ 4\lambda^2+\lambda_c^2 }}   .
$
Note that this implies that~$\frac{\partial}{\partial \lambda} R>0$.
Differentiating again and simplifying yields
$
                \frac{\partial^2}{\partial \lambda^2} R=-2 \lambda_c^2 ( 4\lambda^2+\lambda_c^2 )^{-3/2 } < 0.
$
Similarly,
$
                \frac{\partial}{\partial \lambda_c} R=
                  (1- \frac{ \lambda_c }{ \sqrt{ 4\lambda^2+\lambda_c^2 }}  )/2 >0,
$
$                \frac{\partial^2}{\partial \lambda_c^2} R=-2 \lambda ^2 ( 4\lambda^2+\lambda_c^2 )^{-3/2 } < 0
$,
and
$
                \frac{\partial^2}{\partial \lambda_c \partial \lambda} R= 2\lambda \lambda_c ( 4\lambda^2+\lambda_c^2 )^{-3/2 } > 0 .
$
Thus, the Hessian matrix of~$R$ is
\begin{align}\label{eq:hes_mat}
                        H: &=\begin{bmatrix}  \frac{\partial^2 R}{\partial \lambda^2} &
                           \frac{\partial^2R}{\partial \lambda_c \partial \lambda} \\
                           \frac{\partial^2R}{\partial \lambda_c \partial \lambda} &
                           \frac{\partial^2 R}{\partial \lambda_c^2}
                           \end{bmatrix}\\
                           &=
 \gamma \begin{bmatrix}
                           -  \lambda_c^2  &
                             \lambda \lambda_c  \\
                             \lambda \lambda_c &
                           -  \lambda ^2
                           \end{bmatrix},\nonumber
 \end{align}
where~$\gamma:= 2  ( 4\lambda^2+\lambda_c^2 )^{-3/2 } $.
The eigenvalues of~$H$ are~$0$ and~$-(\lambda^2+\lambda_c^2)\gamma$.
  Recall that a twice differentiable function is a concave function of its
 parameters if and only if its Hessian matrix is negative semidefinite (see  e.g.~\cite{convex_boyd}), i.e. if and only if
 all its eigenvalues are non-positive.
 It follows that
  for~$n=2$
the mapping~$(\lambda,\lambda_c) \to R$
is  {concave}. Fig.~\ref{fig:conc} depicts~$R$ in~\eqref{eq:rfunc}
as a function of its arguments. It may be seen that this is indeed a concave function.
 \begin{figure}[t]
  \begin{center}
  \includegraphics[width= 7cm,height=6cm]{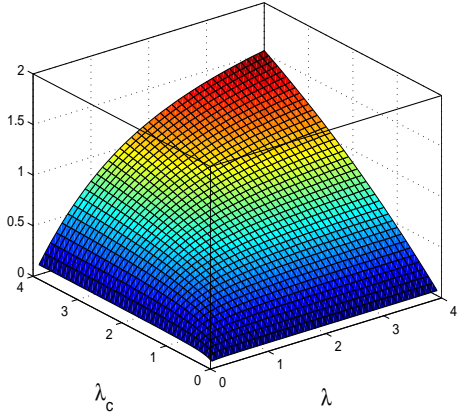}
  \caption{ Steady-state translation rate $R(\lambda,\lambda_c)$ as a function of~$\lambda$ and~$\lambda_c$ for~$n=2$. }\label{fig:conc}
  \end{center}
\end{figure}
\end{Example}

\begin{figure}[t]
  \begin{center}
  \includegraphics[width= 7cm,height=6cm]{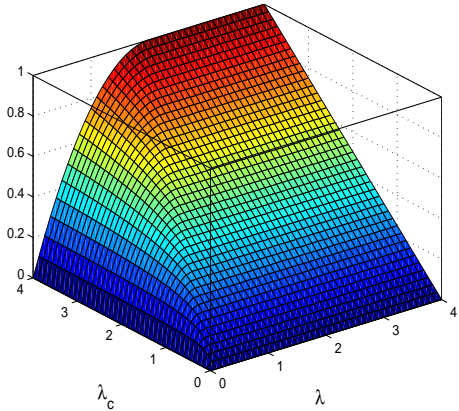}
\caption{Steady-state
 translation rate $\tilde R (\lambda,\lambda_c)$ as a function of~$\lambda$ and~$\lambda_c$.}\label{fig:justadded}
  \end{center}
\end{figure}

\begin{Example}\label{exa:inf}
Consider   the HRFM with~$n\to\infty$, i.e. with the length of the chain going to infinity.
As shown in~\cite{zarai_infi}, in this case~$\tilde  R (\lambda,\lambda_c):=\lim_{n\to \infty}R(\lambda,\lambda_c)$
 exists and satisfies
\be\label{eq:rinf}
            \tilde R (\lambda,\lambda_c)=\begin{cases}
                          \lambda-\lambda^2/\lambda_c, & \lambda <   \lambda_c/2, \\
                           \lambda_c/4, & \lambda \geq \lambda_c/ 2 .
            \end{cases}
\ee
In view of Theorem~\ref{thm:concave}, we expect~$\tilde R $ to be a concave function.
Indeed, this may be observed from Fig.~\ref{fig:justadded} that
depicts~$\tilde R (\lambda,\lambda_c)$ as a function of its variables. This could also be verified analytically from~\eqref{eq:rinf}.
\end{Example}

Recall that a function~$f(\cdot):\R^k_+ \to \R$ is called \emph{positively homogeneous}
if~$f(c x)=cf(x)$ for all~$c>0$ and all~$x\in\R^k_+$.
Since in~\eqref{eq:cf}
$R$ always appears in a term in the form~$R/\lambda_i$,
it follows that~$R$  in the RFM  is positively homogeneous.
In other words,
\be\label{eq:R_is_hom}
            R(c \lambda,c \lambda_1,\dots,c\lambda_n)=c    R(\lambda,\lambda_1,\dots,\lambda_n),\quad\text{for all } c>0.
\ee
From a biophysical point of view this
 means that scaling the initiation rate
and all the transition rates by the same multiplicative  factor $c>0$ in the RFM
 yields an increase of the steady-state translation rate by a factor of~$c$.

Recall that a function~$f(\cdot):\R^k_+ \to \R$ is called \emph{superadditive}
if~$f(x+y)\geq f(x)+f(y)$
  for all~$x,y\in\R^k_+$.
It is well-known that for a positively homogeneous function,
concavity is equivalent to superadditivity
(see, e.g.,~\cite{fund_convx_ana}). Combining this with Theorem~\ref{thm:concave}
 yields the following result.
\begin{Corollary}\label{prop:supadd}
Consider the HRFM with  dimension~$n\geq 2$.
The function~$R=R(\lambda,\lambda_c)$ is superadditive on~$\R^2_+$.
\end{Corollary}
This means that
\[
            R(\lambda+\bar \lambda,\lambda_c+\bar \lambda_c) \geq
            R(\lambda ,\lambda_c )+R(\bar \lambda, \bar \lambda_c),
\]
for all~$\lambda,\lambda_c,\bar \lambda,\bar\lambda_c \geq 0$.
From a biophysical point of view this means the following.
Consider two HRFMs,
 one with
 initiation rate~$\lambda$ and transition rate~$\lambda_c$,
 and the second with initiation rate~$\bar \lambda$ and transition rate~$\bar \lambda_c$.
 The total production rate of these two HRFMs is smaller or equal than
   the production rate of a single HRFM with parameters~$\lambda+\bar\lambda$ and~$\lambda_c+\bar\lambda_c$.

\subsection{Maximizing translation rate}
Consider the problem
of  determining the parameter values~$\lambda_c,\lambda$ that \emph{maximize}~$R$
or, equivalently, that \emph{minimize}~$-R$, in the HRFM.
Obviously, to make this problem meaningful we must constrain the possible parameter values.
This leads to the following optimization problem.
 \begin{Problem}\label{prob:max}
Given the parameters $w_1,w_2, b >0$, minimize~$-R = -R(\lambda ,\lambda_c)$, with respect to its parameters $\lambda$ and $\lambda_c$, subject to the constraints:
\begin{align}\label{eq:constraint}
w_1 \lambda_c+w_2 \lambda  &\leq b, \\
\lambda_c,\lambda &\geq 0.\nonumber
\end{align}
\end{Problem}

In other words, the problem is to maximize the protein translation rate
under an affine  constraint on the total available ``resources'', namely,
the initiation rate~$\lambda$ and the common
transition rate~$\lambda_c$. The constraint on~$\lambda$ [$\lambda_c$] may be related,
among others, to the number of intracellular ribosomes [number of intracellular tRNA molecules].
The values of~$w_1,w_2$ can be used to provide  a
  different weighting  to these two  resources.

Theorem~\ref{thm:concave}
implies that Problem~\ref{prob:max} is a \emph{convex}
 optimization problem~\cite{convex_boyd}. It thus benefits from many desirable properties.
  In particular, it always admits a  solution~$(\lambda^*,\lambda_c^*)$, and
   it can be solved numerically using   efficient algorithms.

The next result shows that increasing $\lambda$ or $\lambda_c$ can only increase the translation rate.
\begin{Proposition}\label{prop:pos_derv}
Consider the HRFM with $n\ge2$. Then
$\frac{\partial R}{\partial \lambda} >0$, and
$\frac{\partial R}{\partial \lambda_c} >0.
$
\end{Proposition}
\begin{Remark}\label{rem:hom}
Note that this implies that the first constraint  in~\eqref{eq:constraint}
can always be replaced by
\be \label{eq:hard_const}
w_1\lambda_c+w_2\lambda = b.
\ee
\end{Remark}

\begin{figure}[t]
  \begin{center}
  \includegraphics[height=6cm]{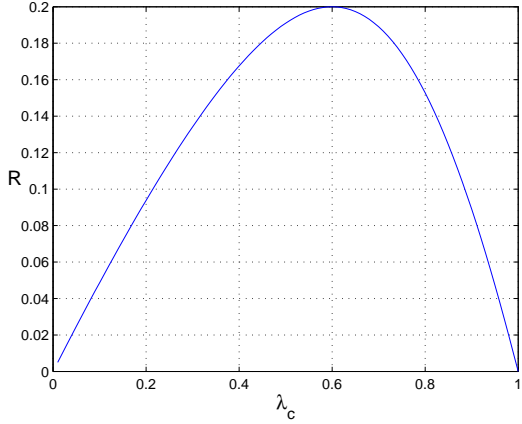}
  \caption{ Translation rate~$R$ as a function of~$\lambda_c$ for the parameters
  in Example~\ref{exa:hrfm2}. }\label{fig:ple2}
  \end{center}
\end{figure}

\begin{Example}\label{exa:hrfm2}
 Consider Problem~\ref{prob:max} for the HRFM with dimension~$n=2$, and
with~$b=w_1=w_2= 1$,  i.e. the constraint   is
\be \label{eq:kigy}
  \lambda_c+   \lambda  = 1.
\ee
Substituting this in~\eqref{eq:rfunc} yields
\[
                R =
               \left  ( 2-\lambda_c-\sqrt{   4(1-\lambda_c)^2 + \lambda_c^2 }  \right    )/2.
\]
Fig.~\ref{fig:ple2} depicts~$R$ as a  function of~$\lambda_c$.
It may be seen that~$R=0$ when~$\lambda_c=0$, as  a zero  transition
rate means zero translation rate,
and also when~$\lambda_c=1$, as then the initiation rate is~$\lambda=1-\lambda_c=0$.
The maximal value,~$R^*=0.2$,   is obtained
 for~$\lambda_c^*=0.6$,
so~$\lambda^*=1-\lambda_c^*=0.4$.
\end{Example}

In general, one cannot expect an \emph{algebraic}
 expression for~$R$ in terms of~$\lambda,\lambda_c$.
 This is true already for the case~$n=2$ (see~\eqref{eq:rfunc}).
 Surprisingly, perhaps, it is possible to give an algebraic
  expression for the optimal
 value~$R^* =R(\lambda^*,\lambda_c^*)$ as a function of the optimal parameter
 values~$\lambda^*,\lambda_c^*$ and the parameters in the affine constraint.
\begin{Theorem}\label{thm:main}
Consider Problem~\ref{prob:max}  for the HRFM with $n\ge2$.
Then
\be \label{eq:mainen}
 R^*= \frac{w_2(n(\lambda^*\lambda_c^*-(\lambda^*)^2)+\lambda^*\lambda_c^*-2(\lambda^*)^2)-w_1(\lambda_c^*)^2}{w_2(\lambda_c^*(n+1)-2\lambda^*)-4w_1\lambda_c^*}.
\ee
\end{Theorem}

\begin{Example}
Consider again Example~\ref{exa:hrfm2}.
Substituting~$n=2$, $w_1=w_2=b=1$, $\lambda^*=0.4$ and~$\lambda_c^*=0.6$ in~\eqref{eq:mainen}
yields
\begin{align*}
R^*   =0.2,
\end{align*}
and this agrees with the result in Example~\ref{exa:hrfm2}.
\end{Example}

When the dimension~$n$ of the HRFM goes to infinity
we can say much more about the optimal solution.
\subsection{Optimizing the infinite-dimensional HRFM}
\begin{Proposition}\label{prop:inf}
Consider Problem~\ref{prob:max} for the infinite-dimensional HRFM.
The optimal values are given by
\begin{align}\label{eq:opt_inf}
                    \tilde \lambda_c^* & = b/\sqrt{w_1(w_1+w_2)},\nonumber\\
\tilde   \lambda^* & =  b\left(   1-\sqrt{ w_1/(w_1+w_2)  }    \right )/w_2  ,\nonumber \\
       \tilde R ^* & =b\left (     2w_1+w_2 -2    \sqrt{w_1(w_1+w_2)}       \right  )  /w_2^2   .
\end{align}
\end{Proposition}

In other words,  for~$n\to \infty$ we have
 simple closed-form expressions for
the solution of Problem~\ref{prob:max} in terms of the constraint
 parameters~$w_1$, $w_2$, and~$b$.

The expression in~\eqref{eq:opt_inf} shows that~$ \tilde R ^*$ increases linearly with~$b$.
This is reasonable, as increasing~$b$ corresponds to allowing larger
values of~$\lambda$ and~$\lambda_c$.

\begin{figure}[t]
  \begin{center}
  \includegraphics[height=6cm]{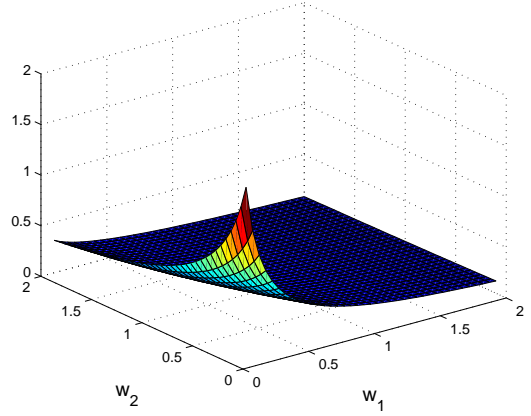}
\caption{$ \tilde             R ^* /b$ in~\eqref{eq:opt_inf} as a function of~$w_1 \in [0.1,2]$ and~$w_2\in[0.1,2]$.}\label{fig:R_opt_infi}
  \end{center}
\end{figure}

Fig.~\ref{fig:R_opt_infi} depicts~$ \tilde             R ^* /b$ as a function of~$w_1$ and~$w_2$.
It may be observed that for large values of either~$w_1$ or~$w_2$
the optimal value~$ \tilde   R ^* /b$ decreases quickly. This is reasonable, as a large value of~$w_1$
[$w_2$] implies a tight constraint on~$\lambda_c$ [$\lambda$], and decreasing any one of these rates implies a small
translation rate. On the other-hand, when both~$w_1$ and~$w_2$ go to zero, $\tilde             R ^* /b$
increases quickly.

Let $\alpha:=w_2/w_1$. Then it follows from \eqref{eq:opt_inf} that
\be \label{eq:opt_inf_l_ratio}
\tilde \lambda_c^* / \tilde \lambda^* = 1+\sqrt{1+\alpha}.
\ee
Thus,  the ratio $\tilde \lambda_c^* / \tilde \lambda^*$
is a strictly increasing and concave function of~$\alpha$.
Note that~\eqref{eq:opt_inf_l_ratio} also
implies that~$\tilde  \lambda_c^* / \tilde \lambda^*\ge2$.
In other words, in the infinite-dimensional HRFM
the optimal  transition rate is always at least twice as big as the
initiation   rate.

The parameters $  \lambda_c $ and $  \lambda $ in the optimization problem
are upper-bounded by $b/w_1$ and $b/w_2$ respectively (see \eqref{eq:constraint}).
It follows from~\eqref{eq:opt_inf} that
\begin{align*}
                    \lim_{w_1 \to 0} \begin{bmatrix}\tilde \lambda_c^* ,&  \tilde \lambda^*, & \tilde R^*\end{bmatrix}&=
                     \begin{bmatrix} \infty ,& b/w_2  ,&  b/w_2 \end{bmatrix}, \\
                    \lim_{w_2 \to 0} \begin{bmatrix}\tilde \lambda_c^*, &  \tilde \lambda^* ,& \tilde R^*\end{bmatrix}&=
                     \begin{bmatrix} b/w_1, & b/(2w_1) , &  b/(4w_1) \end{bmatrix}.
\end{align*}
The case~$w_1\to 0$ implies that there is no constraint on~$\lambda_c$ and thus~$ \tilde \lambda_c^*=\infty$.
Also, the maximal possible value for~$\lambda$ is~$ \tilde \lambda^*=b/w_2$. This becomes the rate limiting factor so~$\tilde R^*= \tilde
\lambda^* = b/w_2$.
When~$w_2\to 0$ the constraint~\eqref{eq:hard_const} yields~$\tilde \lambda_c^*=b/w_1 $. Also, in this case~$\alpha=0$
and the ratio in~\eqref{eq:opt_inf_l_ratio} attains its minimal value, namely,
  $\tilde \lambda_c^* /  \tilde \lambda^* =2 $, so $\tilde \lambda^*=\tilde \lambda_c^* / 2  = b/(2w_1)$.

It turns out that the  expressions in~\eqref{eq:opt_inf}
actually provide good approximations for the
optimal  parameter  values in  \emph{finite-dimensional} HRFMs.
The next example demonstrates this.
\begin{Example}\label{exa:n20}
Consider Problem~\ref{prob:max}
for the HRFM with~$n=20$, and~$w_1=w_2=b=1$.
Applying a simple numerical algorithm to solve Problem~\ref{prob:max}
yields
\[
\lambda^*_c= 0.7069,\; \lambda^*=0.2931,\;   R^*=0.1716,
\]
(all numbers are to four digit accuracy).
On the other-hand, for~$w_1=w_2=b=1$ Eq.~\eqref{eq:opt_inf} yields
\begin{align*}
                    \tilde \lambda_c^*  & = 1/ \sqrt{2} \approx 0.7071  , \\
                    \tilde \lambda^*   &=  1-\sqrt{1/2}\approx  0.2929  , \\
                    \tilde R ^*   &=       3-2\sqrt{2} \approx 0.1716  .
\end{align*}
  Thus, the optimal values for the infinite-dimensional HRFM agree well
  with the optimal values already for the $20$-dimensional HRFM.
\end{Example}

It is important to note
 that the typical length of mRNA sequences is
larger than~$20$ sites. For example, in   {\em S. cerevisiae} the mean length
 is about~$33$  sites; in mammals  the mRNA chains
  are much longer; thus,
 the closed-form asymptotic results here
   provide a good approximation for the optimal parameter values
   in finite-dimensional  HRFM models of gene translation.


 \section{A biological example}\label{sec:bio}
There exist  effective
 experimental
 approaches for estimating  the
 translation-elongation rates and
 the protein synthesis  rate, but currently
  there is no  experimental  approach for measuring
 the  initiation rate.
  Indeed, initiation is a highly complex mechanism and its efficiency  is based on numerous
  biophysical
   properties of the coding sequence including:
    the nucleotide context of the START codon (i.e., the first codon that is translated in a gene) \cite{Kozak1986,Zur2013};
     the folding of the RNA near the beginning of the open reading frame~(ORF) and the nucleotide composition in this region \cite{Zur2013,Tuller2010b};
      the number of ribosomes and mRNA molecules in the cell;
       the length and the nucleotide context of the 5'UTR; interaction between initiation and elongation steps \cite{Zur2013,Tuller2010b},
        and more. Thus, although there exist
         experimental approaches for measuring positions on the mRNA suspected to   correspond
          to   initiation sites~\cite{Ingolia2011,Lee2012}, there are   no
           large scale direct measurements of
           initiation rate.

           Several papers addressed the problem
           of estimating  the initiation rate using computational models
           of translation~\cite{zarai_infi,Salis_opt_rbs,10.1371/journal.pcbi.1002866}.
           One possible application of our results is to estimate the
             initiation rate  based on measurements of elongation   and translation rates.
             Indeed,   we may assume, without loss of generality, that~$b=1$.
              Then,  given~$\tilde R^*$ and~$\tilde \lambda^*_c$,   we can determine~$w_1,w_2$
             based on~\eqref{eq:opt_inf}. Plugging~$w_1,w_2$   back in~\eqref{eq:opt_inf}
              yields the initiation  rate~$\tilde \lambda^*$.
             The underlying  assumptions here are that  the mRNA chain
             is relatively long; that all elongation rates are equal;
               and that the parameters of the translation process are indeed optimized by
             evolution.

             To demonstrate this we consider a specific example.
Ingolia {\em et al.} \cite{Ingolia2011} estimated the constant
 transition  rate in {\em M. musculus} embryonic stem cell
  by applying cyclohexamide to halt translation, and harringtonin  to halt   initiation
  at different time steps. This allows   estimating the speed of elongation by
    measuring the movement of the ``ribosomal density wave''. They
concluded  that   in mouse embryonic cells $5.6$ codons are translated per second (in terms
of the HRFM, this   corresponds
  to $\lambda_c=5.6/15=0.3733$ sites per second (all numbers are to four digit accuracy), as a ribosome spans
   about~$15$ codons~\cite{Ingolia2009}). According to~\cite{Ingolia2011}, this elongation speed is  typical,
   and   does not vary much  between different genes.

A recent study by Schwanhausser {\em et al.}~\cite{Schwanhausser2011}
 estimated the translation rate in {\em M. musculus} fibroblasts by simultaneously measuring protein abundance and turnover by parallel metabolic pulse labeling for more than $5000$ genes in mouse. They found that the
median translation rate in mouse is about~$40$ proteins per mRNA per hour (i.e.,
  $R=40/3600= 0.0111$ proteins per mRNA per second).

Summarizing, in terms of the HRFM these biological findings suggest that $\lambda_c^*= 0.3733$ and~$R^*=0.0111 $.
To estimate the initiation rate~$\lambda^*$ in mouse, we plug
these values (and~$b=1$) in~\eqref{eq:opt_inf}.
This yields~$(w_1,w_2)= (84.6477,-84.5629 )$ or~$(w_1,w_2)= (0.0848, 84.5629 )$. The first case is
impossible, as in our optimization problem
the~$w_i$s must be positive, so we conclude that these
values correspond to a solution of Problem~\ref{prob:max} with
$ (w_1,w_2,b)= (0.0848, 84.5629, 1 ) $.
Applying~\eqref{eq:opt_inf} with these values yields
 $\lambda^*= 0.0114$ sites per second. This corresponds  to~$0.1718$ codons per second.
Note that this agrees well with the estimate in~\cite{zarai_infi}
  that was obtained  using~\eqref{eq:rinf}. However,  the approach here is based on a different
  argument, namely, that evolution shaped  the parameters of the translation process
  so that
  they correspond to an optimal solution for Problem~\ref{prob:max}.

\section{Discussion and Conclusion}
The RFM is a   deterministic computational model for ribosome flow along the mRNA.
 It may be viewed as a mean-field approximation of the stochastic TASEP model
 and in particular encapsulates both the simple exclusion and the total asymmetry properties of
    TASEP.

   The RFM is characterized by an order $n$, corresponding to the number of sites along the mRNA chain,
   a positive initiation rate~$\lambda$ and a set of positive transition rates~$\lambda_1,\dots,\lambda_n$.
Under the assumption (or approximation) of equal transition rates (i.e. $\lambda_1=\dots=\lambda_n:=\lambda_c$),
the RFM becomes the HRFM.
Recent studies have suggested that this is the case in some organisms/conditions \cite{Ingolia2011,Qian2012}.

 In this paper, we showed   that in the HRFM the
  steady-state
  translation rate~$R=R(\lambda,\lambda_c)$ is a \emph{concave} function of its parameters.
   This implies that a local maximum of~$R$ is the global maximum.
  Furthermore, this allows posing
   the problem of maximizing the steady-state translation rate~$R$
  in a meaningful way as a convex optimization problem. Such
   problems
   can be solved using efficient numerical algorithms  (see, e.g.,~\cite{Dimitri1999,Kiwiel1985,Bonnans2006}).

  We also provide an explicit algebraic expression for the optimal
  translation
  rate~$R^*$  in terms of
  the  optimal parameter values~$\lambda^*,\lambda_c^*$,
  and the parameters in the affine constraint~$w_1,w_2,$ and~$b$,
  as well as an \emph{explicit} solution to the convex optimization problem in  the
  infinite-dimensional HRFM.

The reported results may potentially
 be
 used for re-engineering gene expression for various biotechnological applications.
  Specifically, an important problem is to optimize
   the translation efficiency and protein levels of heterologous genes in a new host \cite{Plotkin2011,Tuller2010,Gustafsson2004,Kimchi-Sarfaty2013}.
 In Section~\ref{sec:bio} we show how $w_1$ and $w_2$ can be estimated. The idea is to use the explicit equations for $R^*$, $\lambda^*$ and $\lambda_c^*$ in the infinite-dimensional HRFM. We show that based on experimental measurements of the elongation rates ($\lambda_c^*$), and translation rates ($R^*$), we can estimate $w_1$ and $w_2$. The underlying assumptions for this approach are that: (1) evolution optimized the translation process; and (2) the mRNA chain is relatively long.
In addition, we would like to emphasize that in the case of biotechnological engineering of gene translation, $w_1$ and $w_2$ may be evaluated based on intra-cellular measurement of the concentration and metabolic costs of proteins and genes related to the translation machinery such as initiation factors, elongation factors, tRNA molecules, aminoacyl-tRNA synthetases, etc; these values are related to the 'cost' of increasing $\lambda$ and $\lambda_c$.

Our results may also be related to  the
  evolution of gene expression,  and specifically translation and transcript sequences.
   Indeed, translation is the process consuming most of the cell energy~\cite{Plotkin2011,Tuller2010,Alberts2002},
 and  it is reasonable to assume that for organisms under strong evolutionary pressure, evolution shapes
 the genomic machinery
  so that it optimizes the protein translation rate for the given finite resources.

A natural  topic for further  research
is  whether  the steady-state  translation
rate~$ R=R(\lambda,\lambda_1,\dots,\lambda_n)$ in the~RFM  is
a concave function of its parameters.
This question seems to be technically more demanding,
as the Hessian matrix of the~$n$-dimensional RFM
has dimensions~$(n+1) \times (n+1)$, whereas that of the HRFM is~$2 \times 2$.
A more general question is related to the concavity of other
 models of translation   including
  various versions of the TASEP model \cite{RFM_feedback,TullerGB2011,Tuller2010,Shah2013,Chu2012,Potapov2012}.

More generally,   the asymmetric simple exclusion
process~(ASEP) has become the ``default stochastic model for transport phenomena''
\cite{Yau2004}, and has been used to model and analyze many important
natural and artificial
processes~\cite{TASEP_book}.  We believe that the~RFM and the HRFM
 can also be applied to
  model and analyze more natural and artificial
processes.

\section*{Acknowledgements}
We thank the anonymous reviewers for their  helpful comments.

\section*{Appendix -- Proofs}\label{sec:proofs}
 {\sl Proof of Theorem~\ref{thm:concave}.}
The proof is based on analyzing the Hessian matrix of~$R$.
  Define the \emph{normalized  initiation  rate}~$\eta$ by
\[
            \eta:=\lambda/\lambda_c.
\]
Then we can rewrite~\eqref{eq:cfper} as
\be\label{eq:eetf}
                {e_n}  ={\eta}f_n(e_n),
\ee
with
\be\label{eq:deffn}
                        f_n(z):=1- \cfrac{z          }
                                  {  1-\cfrac{z}
                                  { 1-\cfrac{z }{\hphantom{aaaaaaa} \ddots
                             \genfrac{}{}{0pt}{0}{}
                             {1-\cfrac{z}{1-z  }} }}},
\ee
where on the right-hand side $z$ appears~$n$ times.
Note that~$f_n(z)$ is not necessarily well-defined for all~$z \in (0,1)$.
For example,
\begin{align*}
                        f_3(z)&= 1- \cfrac{z}{1-\cfrac{z}{1-z }}\\
                                  &=\frac{z^2-3z+1}{-2z+1}
\end{align*}
is not well-defined for~$z=1/2$.

Eq.~\eqref{eq:eetf} implies that~$e_n=e_n(\eta)$,
and since~$e_n\in(0,1)$,
\[
                0<f_n(e_n)<1/\eta.
\]

The following results are needed to prove Theorem~\ref{thm:concave}.
\begin{Proposition}\label{prop:fff}
Fix arbitrary~$n\geq 2$ and~$\eta>0$. Let~$e_n=e_n(\eta)$.
Then for all~$z \in[0,e_n]$ the functions~$f_n(z),f'_n(z):=df_n(z)/dz$,
 and~$f''_n(z):=d^2f_n(z)/dz^2$ are well-defined and  satisfy
\begin{align}\label{eq:fff}
           f_n(z)&>0,  \nonumber\\
           f'_n(z)&<0, \nonumber \\
           f''_n(z)&<0 .
\end{align}
\end{Proposition}

\begin{figure}[t]
  \begin{center}
  \includegraphics[width=9cm,height=7cm]{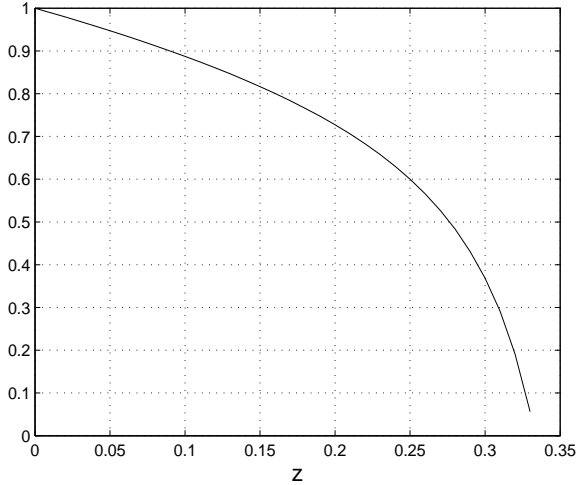}
  \caption{ $f_4(z)$ as a function of~$z$.
  }\label{fig:fpp}
  \end{center}
\end{figure}

In other words, for all~$z \in[0,e_n]$ the function~$f_n(z)$
is positive, strictly decreasing, and concave.
\begin{Example}
Consider the case~$n=4$.
It follows from the results in~\cite{HRFM_steady_state} that  for all~$\eta>0$,
$ e_4(\eta) \in (0, a)$,
where~$a:=1/(4  \cos^{ 2}(\pi/6)) =1/3$.
Fig.~\ref{fig:fpp}
depicts the function~$f_4(z)$
for~$z \in [0,a ]$.
It may be seen that~$f_4(z)$ is well-defined,
positive, strictly decreasing, and concave in this range.
\end{Example}

 {\sl Proof of Proposition~\ref{prop:fff}.}
Pick~$\eta>0$. The proof is by induction on~$n$. For~$n=2$,~$f_2(z)=1-\frac{z}{1-z}$, so~$f_2(z)>0$ for all~$z \in[0,1/2)$.
By~\eqref{eq:e2here},
\[
            e_2=\eta+(1/2)-\sqrt{\eta^2+1/4},
\]
so~$e_2<1/2$.
Differentiating~$f_2$ yields
\begin{align*}
                        f_2'(z)&=-(1-z)^{-2},\\
                        f_2''(z)&=-2(1-z)^{-3}.
\end{align*}
Thus for~$n=2$, Eq.~\eqref{eq:fff} holds for all~$z \in[0,e_2]$.

 For the induction step,
it is useful to let~$p \in (0,1)^n$ denote the equilibrium point of
 the~$n$-dimensional HRFM,
and let~$q \in (0,1)^{n+1}$ denote  the equilibrium point of the~$(n+1)$-dimensional
 HRFM.
It was shown in \cite[Proposition~1]{zarai_infi} that
 \be \label{eq:cpq}
                            q_{n+1}<p_n.
 \ee
In other words,  for two HRFM chains
  with the same~$\eta$ the translation rate in the longer chain is smaller
 than the translation rate in the shorter chain.
  Assume that~\eqref{eq:fff} holds for all~$z \in [0,p_n]$.
By~\eqref{eq:deffn},
\be\label{eq:deffnp1}
                f_{n+1}(z)= 1-\frac{z}{f_n(z)}.
\ee
By the induction hypothesis,~$f_n(z)>0$ for all~$z \in [0,p_n]$, so~$f_{n+1}(z)$ is well-defined for all~$z \in [0,p_{n }]$.
Differentiating~\eqref{eq:deffnp1} yields
\begin{align}\label{eq:f'np1}
                        f'_{n+1}(z)&=\frac{f_n'(z)z-f_n(z)}{ f_n^2(z) }.
\end{align}
 Combining this with the induction hypothesis  implies that the right-hand side
of~\eqref{eq:f'np1} is well-defined and strictly negative for all~$z \in [0,p_n]$,  so
in particular
\be\label{eq:f'isdone}
                        f'_{n+1}(z) <0, \quad \text{for all } z\in[0,q_{n+1}].
\ee
We now show that
\be\label{eq:fnp1zp}
f_{n+1}(z)>0, \quad \text{for all }z \in [0,q_{n+1}].
\ee
Seeking a contradiction, assume that~\eqref{eq:fnp1zp} does not hold. Then since~$f_{n+1}(0)=1$,  there exists a minimal~$y \in [0,q_{n+1}]$
such that~$f_{n+1}(y)=0$. Combining this with~\eqref{eq:f'isdone} implies that
$ f_{n+1}(q_{n+1})\leq 0  $. But since~$q\in (0,1)^{n+1}$ is the equilibrium point of the~$(n+1)$-dimensional HRFM,
$
                q_{n+1}=\eta f_{n+1}(q_{n+1})
$, so~$q_{n+1}\leq 0$.
This contradiction proves~\eqref{eq:fnp1zp}.

Differentiating~\eqref{eq:f'np1} yields
\[
                        f''_{n+1}(z ) =\frac{ zf''_n(z) f^2_n(z)-2zf_n(z)(f'_n(z))^2+2f^2_n(z)f'_n(z)}
                                      { f_n^4(z) },
\]
  and using~\eqref{eq:f'np1} yields
\begin{align}\label{eq:f2np1}
                        f''_{n+1}(z)
                        &= \frac{ zf''_n(z) f^2_n(z)  -2 f_n^3(z) f'_n(z)f'_{n+1}(z) }{ f_n^4(z) }.
\end{align}
We already know that~$f'_{n+1}(z)<0$ for all~$z\in[0,p_n]$ and combining this
with the induction hypothesis implies that
\be\label{eq:f2df}
                    f''_{n+1}(z)<0,\quad \text{for all } z \in [0,p_n].
\ee
Combining~\eqref{eq:fnp1zp}, \eqref{eq:f'isdone}, \eqref{eq:f2df}, and~\eqref{eq:cpq} completes the proof of the induction step.~$\square$

\begin{Proposition}\label{prop:hprime}
Fix an arbitrary~$n\geq 2$. Let~$h_n(\cdot):\R_+ \to (0,1) $ be the function such that~$e_n=h_n(\eta)$.
Then
\begin{align}\label{eq:pott}
h_n'(\eta  ):=\frac{dh_n(\eta)}{d\eta} &>0, \nonumber \\
h_n''(\eta  ):=\frac{d^2h_n(\eta)}{d\eta^2}&<0,
\end{align}
for all~$\eta>0$.
\end{Proposition}

In other words,
the mapping~$\eta \to e_n(\eta)$ is   strictly  increasing
and concave.

{\sl Proof of Proposition~\ref{prop:hprime}.}
We can  write~\eqref{eq:eetf} as a polynomial equation in~$e_n$
with coefficients that are smooth functions of~$\eta$. It is possible
to show that the feasible~$e_n$ (i.e. the one corresponding to the solution~$e \in Int(C)$)
is a simple root of this polynomial for all~$\eta>0$~\cite{sensi_RFM}.
Hence,~$h_n(\cdot)$ is a smooth function.

Rewriting~\eqref{eq:eetf} as~$h_n(\eta) = \eta f_n(h_n(\eta)) $ and differentiating with respect to~$\eta$ yields
\be\label{eq:hdot}
 \left(1-\eta f'_n(h_n)\right) h'_n =f_n(h_n).
\ee
Combining this with Proposition~\ref{prop:fff} implies that~$h'_n>0$.
Differentiating~\eqref{eq:hdot} with respect to~$\eta$ yields
\[
\left(1-\eta f'_n(h_n)\right) h_n'' =2h'_n f'_n(h_n)+\eta f_n''(h_n)(h'_n)^2.
\]
Combining this with the fact that~$h'_n(\eta)>0$
 and Proposition~\ref{prop:fff}  implies that~$h_n''(\eta)<0$,
   and this completes the proof of Proposition~\ref{prop:hprime}.~$\square$

We can now complete the proof
 of Theorem~\ref{thm:concave}.
Differentiating~$R=\lambda_c h(\eta)$ with respect to~$\lambda_c$ yields
\begin{align}\label{eq:rlamc}
                        \frac{\partial R}{\partial \lambda_c}= h_n(\eta)- \lambda
                          \lambda_c^{-1}h'_n(\eta),
\end{align}
and
\begin{align*}
\frac{\partial^2 R}{\partial \lambda_c^2}
   &=\lambda^2 \lambda_c^{-3} h_n''(\eta).
\end{align*}
Similarly,
\begin{align}\label{eq:rparlam}
\frac{\partial  R}{\partial \lambda }
   &=h_n'(\eta),\\
\frac{\partial^2  R}{\partial \lambda \partial \lambda_c }
   &=  -\lambda \lambda_c^{-2} h''_n(\eta)  ,\nonumber \\
\frac{\partial^2  R}{\partial \lambda^2 }
   &=   \lambda_c^{-1} h''_n(\eta) \nonumber .
\end{align}
Substituting these expressions in the Hessian  matrix~\eqref{eq:hes_mat}
yields
\[
                        H =
                        \begin{bmatrix}  \lambda_c^{-1} &
                           -\lambda \lambda_c^{-2} \\
                           -\lambda \lambda_c^{-2} &
                           \lambda^2 \lambda_c^{-3}
                           \end{bmatrix} h_n''(\eta),
\]
A calculation shows that the eigenvalues of~$H$ are~$0$ and~$  ( \lambda_c^{2} +\lambda^2 )\lambda_c^{-3} h''(\eta)$. Since~$h_n''(\eta)<0$
this implies that~$H$ is a negative semidefinite matrix. This
completes the proof of Theorem~\ref{thm:concave}.~$\square$

{\sl Proof of Proposition~\ref{prop:pos_derv}.}
Since~$f'_n(e_n)<0$,
it follows from~\eqref{eq:hdot} that~$h_n'(\eta)<f_n(h_n(\eta))$, so~\eqref{eq:eetf}
 yields~$h_n'(\eta) <e_n/\eta=h_n(\eta)/\eta$. Combining this with~\eqref{eq:rlamc}
implies that
\[ \partial R / \partial \lambda_c>0.\]

Using \eqref{eq:rparlam} and
Proposition~\ref{prop:hprime} implies that \[\partial R / \partial \lambda>0,\]
and this completes the proof of Proposition~\ref{prop:pos_derv}.~$\square$


{\sl Proof of Theorem~\ref{thm:main}.}
The proof is based on formulating the Lagrangian function associated with Problem~\ref{prob:max}
and determining the optimal parameters by
 equating its   derivatives to zero (see, e.g.,~\cite{convex_boyd}).
We require the following result.
\begin{Proposition}\label{prop:detr}
Consider the~$n$-dimensional HRFM with~$n\geq 2$. Then
\begin{align}
                         \frac{\partial}{\partial \lambda} e_n&=
                      \begin{cases}
                         \frac{3 }{4\lambda (n+3)}   ,& e_n=1/4,\\
                        {\lambda_c}  ( 4 e_n -1 ) /(r \lambda)  , & \text{otherwise},
                      \end{cases} \label{eq:enlam1}\\
                                      \frac{\partial}{\partial \lambda_c} e_n&=
                                      \begin{cases}
                       \frac{-3}{4\lambda_c(n+3)},& e_n=1/4,\\
                       (1-    4e_n)/r , & \text{otherwise},
                                      \end{cases}\label{eq:enlam2}  ,
\end{align}
where
\be\label{eq:defrs}
r   : =   2\lambda_c+  {\lambda_c^2} (n+1) \lambda^{-1}   + ( n+2) (  \lambda-\lambda_c )  e_n^{-1}       .
\ee
\end{Proposition}
{\sl Proof of {Proposition}~\ref{prop:detr}.}
It is well-known that there is a strong connection between
continued fractions and  \emph{Chebyshev polynomials} (see, e.g.,~\cite{mans_2000}).
We begin by stating some properties of  these polynomials
 that are used in the proof. For more details, see e.g.~\cite{cheb_poly}.

The Chebyshev polynomial  of the second kind of degree~$n$
is defined by $U_n(x):=\frac{ \sin{(n+1)\theta}} {\sin{\theta}}$, where $x=\cos{\theta}$.
For example,
\begin{align*}
            U_2(x)&= \frac{ \sin{3\theta}} {\sin{\theta}} \\&= \frac{  -4\sin^3 \theta+3\sin \theta} {\sin{\theta}}\\
            &=-4\sin^2 \theta+3\\&=-4(1-x^2)+3 .
\end{align*}
These polynomials can also be defined recursively by
\begin{align}\label{eq:chv1}
U_0(x)&=1, \nonumber \\
 U_1(x)&=2 x , \nonumber \\
U_{n+1}(x)&=2 x U_n(x)-U_{n-1}(x), \quad n =  1,2,....
\end{align}
It is not difficult to prove
 that this implies
that
\[
U_n(1)=n+1,\quad   U'_n(1)=n(n+1)(n+2)/3
\]
  for all~$n$.
More generally, it is well-known that the  derivative of $U_n(x)$ satisfies
\be \label{eq:un_diff}
2(1-x^2) U_n'(x) = {-nU_{n+1}(x) + (n+2)U_{n-1}(x)} .
\ee

It has been shown in~\cite{RFM_feedback}
that the last coordinate of the equilibrium point
of the $n$-dimensional HRFM satisfies
\be \label{eq:un_en}
\lambda U_{n+1}(s)=\lambda_c U_n(s)e_n^{1/2},
\ee
where
\[
                s:=1/(2\sqrt{e_n}).
\]
%
Note that since~$e_n \in (0,1)$, $s \in (1/2,\infty) $.

Eq.~\eqref{eq:un_en} implies in particular
that
\be\label{eq:unnz}
U_n(s) \ne 0,\quad \text{for all }\lambda,\lambda_c>0.
\ee
Indeed, if~$U_n(s)=0$ then~\eqref{eq:un_en} yields~$U_{n+1}(s)=0$
and then repeatedly applying the  recursive definition~\eqref{eq:chv1}
 yields~$U_0(s)=0$ which is a contradiction.

Differentiating \eqref{eq:un_en} with respect to~$\lambda$ yields
\begin{align*}
    \lambda_c & \left ( \frac{U_n(s)}{2} e_n^{-1/2} \frac{\partial}{\partial \lambda} e_n +
     e_n^{1/2} \frac{\partial}{\partial \lambda}U_n(s) \right  )  \\
      & = U_{n+1}(s)+\lambda \frac{\partial}{\partial \lambda} U_{n+1}(s)\\
      & = \frac{\lambda_c}{ \lambda} U_{n}(s)  e_n^{1/2}    +\lambda \frac{\partial}{\partial \lambda} U_{n+1}(s).
\end{align*}
Thus,
\begin{align}\label{eq:lkiop}
                        \lambda_c & \left(  \frac{1}{2} e_n^{-1/2}  U_n(s)  \frac{\partial}{\partial \lambda} e_n +
                        e_n^{1/2} U'_n(s) \frac{\partial}{\partial \lambda}s \right ) \nonumber \\
                        &=\frac{\lambda_c}{\lambda}  U_{n}(s)  e_n^{1/2}
                        +\lambda  U'_{n+1}(s)  \frac{\partial}{\partial \lambda}s.
\end{align}
By the definition of~$s$,
\[
                    \frac{\partial}{\partial \lambda}s=-\frac{1}{4} e_n^{-3/2} \frac{\partial}{\partial \lambda}e_n,
\]
and substituting this in~\eqref{eq:lkiop}   yields
\be \label{eq:derlam}
                 g   \frac{\partial}{\partial \lambda} e_n=    \frac{4 \lambda_c}{ \lambda   } U_{n}(s)  e_n,
\ee
where
\begin{align}\label{eq:comg}
                g  :
                =2\lambda_c U_n(s)+e_n^{-1}
                 \left( \lambda U'_{n+1}(s)   -\lambda_c e_n^{1/2} U'_{n }(s) \right) .
\end{align}

Differentiating~\eqref{eq:un_en} with respect to~$\lambda_c$ yields
\begin{align}
                      (U_n(s)+\lambda_c   \frac{\partial}{\partial \lambda_c} U_n(s)   )e_n^{1/2}&
                      +\frac{1}{2} \lambda_c U_n(s)   e_n^{-1/2} \frac{\partial}{\partial \lambda_c} e_n \nonumber \\
                      &= \lambda \frac{\partial}{\partial \lambda_c} U_{n+1}(s),\nonumber
\end{align}
and simplifying similarly yields
\be \label{eq:derlamCC}
               g    \frac{\partial}{\partial \lambda_c} e_n=    {-4 U_{n}(s) e_n } .
\ee

We consider two cases.

{\sl Case 1.} Suppose that~$s=1$. Then~$e_n=1/4$
and~\eqref{eq:un_en} yields
$\lambda_c=2 \lambda (n+2)/(n+1)$. Substituting these values in~\eqref{eq:comg} yields
$g= 8 \lambda (n^2+5n+6) /3$. Substituting this in~\eqref{eq:derlam} and~\eqref{eq:derlamCC}
proves Proposition~\ref{prop:detr} in the case~$e_n=1/4$.

{\sl Case 2.} Suppose that~$s\ne 1$ (so~$e_n \ne 1/4$).
To simplify~$g $,
let~$y := \lambda U'_{n+1}(s)-\lambda_c e_n^{1/2} U'_{n }(s) $.
Using \eqref{eq:un_diff} yields
\begin{align*}
                    2(1-s^2)y =&\:\lambda(n+3)U_n(s)-\lambda(n+1) U_{n+2}(s) \nonumber \\
                    &{-}\:\lambda_c(n+2)e_n^{1/2}U_{n-1}(s) +\lambda_c e_n^{1/2} n U_{n+1}(s),\nonumber
\end{align*}
and applying~\eqref{eq:chv1} yields
\begin{align*}
                    2(1-s^2)y &=\lambda(n+3)U_n(s)   \\
                    &- \lambda(n+1) ( 2sU_{n+1}(s)-U_n(s)  )\\
                    &-\lambda_c(n+2)e_n^{1/2}( 2sU_n(s)-U_{n+1}(s)  ) \\
                    &+\lambda_c e_n^{1/2} n U_{n+1}(s)\\
                    &=(2n+4) ( \lambda - s\lambda_c e_n^{1/2}      )U_n(s)\\
                    &+ (2n+2)( - \lambda  s+ \lambda_c e_n^{1/2}    )U_{n+1}(s).
\end{align*}
Substituting~$s=1/(2\sqrt{e_n})$,~$U_{n+1}(s)$ from~\eqref{eq:un_en},
and simplifying yields
\begin{align*}
                2 & (1-s^2)y  \\&=\left( \lambda(2n+4)-\lambda_c (2n+3)+\frac{\lambda_c^2}{\lambda}  (2n+2)e_n   \right )U_n(s).
\end{align*}
Thus,
\begin{align*}
              2 & (1-s^2)      g \\&  =2(1-s^2) \left( 2\lambda_c U_n (s) +e_n^{-1}  y \right) \\
                     &= 4(1-s^2) \lambda_c U_n(s)\\
                     &+ e_n^{-1}   \left( \lambda(2n+4)-\lambda_c (2n+3)+\frac{\lambda_c^2}{\lambda}  (2n+2)e_n  \right )   U_n(s)  ,
\end{align*}
and simplifying this yields
\be\label{eq:simpg}
          (1-e_n^{-1}/4 )    g  = U_n(s) r.
\ee
Substituting \eqref{eq:simpg} in~\eqref{eq:derlam} and~\eqref{eq:derlamCC}
completes the proof of Proposition~\ref{prop:detr}.~$\square$

We can now prove  Theorem~\ref{thm:main}.
The Lagrangian function associated with Problem~\ref{prob:max} is
\[
                L(\lambda_c,\lambda,\theta) := \lambda_c e_n+ (b-w_1\lambda_c-w_2\lambda)\theta,
\]
where $\theta$ is the Lagrange multiplier.
Differentiating  this with respect to~$\lambda_c$ and equating to zero yields
\be\label{eq:jadded}
                        e_n^*+\lambda_c^* \frac{\partial}{\partial \lambda_c}e_n^*=w_1 \theta^*,
\ee
where~$\lambda_c^*,\lambda^*$ are the optimal values of~$\lambda,\lambda_c$,
$e_n^*=e_n(\lambda_c^*,\lambda^*)$ and~$ \frac{\partial}{\partial \lambda_c}e_n^*=\frac{\partial}{\partial \lambda_c}e_n(\lambda_c^*,\lambda^*) $.
Differentiating~$L$ with respect to~$\lambda $ and equating to zero yields
\[
                        \lambda_c^* \frac{\partial}{\partial \lambda }e_n^*=w_2\theta^*,
\]
and combining this with~\eqref{eq:jadded} yields
\be\label{eq:dldc}
           \frac{e_n^*}{\lambda_c^*}=  \frac{w_1}{w_2} \frac{\partial}{\partial \lambda }e_n^*
           -   \frac{\partial}{\partial \lambda_c}e_n^*.
\ee

We now consider two cases.

{\sl Case 1.} Suppose that~$s^*=1$ (i.e.~$e_n^*=1/4)$.
  We know that in this case
  $ \lambda_c^*=2\lambda^*(n+2)/(n+1) $.
  It is straightforward to show that this equation implies that
  the term on the right-hand side of~\eqref{eq:mainen} is~$\lambda_c^*/4$.
  On the other-hand,~$R^*=\lambda_c^* e_n^*=\lambda_c^*/4$.
  This proves~\eqref{eq:mainen} for the case~$s^*=1$.

{\sl Case 2.}    Suppose that~$s^* \ne 1$ (i.e.~$e_n^*  \ne 1/4)$.
Combining~\eqref{eq:dldc}
with~\eqref{eq:enlam1} and~\eqref{eq:enlam2}
\begin{align*}
                             r \frac{e_n}{\lambda_c}= ( 4e_n - 1  )   \frac{b}{  \lambda w_2   } \vert^* ,
\end{align*}
where~$\vert^*$ means that this equation holds for the optimal parameter values.
Substituting~$r $ from~\eqref{eq:defrs} and simplifying yields
\begin{align}\label{eq:aben}
          \alpha^*  e_n^*
          &=  \beta^*  ,
\end{align}
where~$\alpha^* :=   2 \lambda^* \lambda_c^* w_2 +
                     (\lambda_c^*)^2 (n+1)w_2 -4 b \lambda_c^*     $,
 and~$\beta^* :=  (n+2)(\lambda^*_c-\lambda^*) \lambda w_2-b \lambda_c  $.

Suppose for a moment that~$\alpha^*=0$. Then~\eqref{eq:aben} implies that also~$\beta^*=0$ and combining
this with~$b=w_1 \lambda_c^* +w_2 \lambda^*  $ yields
\be\label{eq:wpmat}
                            (nw_2-4 w_1) ( w_2+(n+2) (n w_2-4 w_1)   ) =0.
\ee
This implies that~$\alpha^*=0$ only when~$w_2=(n+2)(4 w_1-n w_2)$, i.e.
$  w_2 = \frac{4 w_1(n+2)}{(n+1)^2}   $.
Thus,~$b=w_1(\lambda_c^*+\frac{4\lambda^* (n+2)}{(n+1)^2})$,
and substituting this in~$\alpha^*=0$ yields
\[
        \lambda^*_c(n+1) =2 \lambda^* (n+2).
\]
                We already know that this corresponds to
                the case~$s=1$, and since we are considering the case~$s \not =1$,
                $\alpha^*\ne 0$, then
\[
            e_n^*=\beta^*/ \alpha^*.
\]
Using the fact that~$b=w_1 \lambda_c^*+w_2 \lambda^*$ completes the proof
of Theorem~\ref{thm:main}.~$\square$

{\sl Proof of Proposition~\ref{prop:inf}.}  Recall that in the infinite-dimensional~HRFM
  the steady-state translation rate~$\tilde R$
is given by~\eqref{eq:rinf}.
We know that the optimal values satisfy~$ w_1 \lambda_c +w_2 \lambda =b $,
so~$ \lambda = (b-w_1 \lambda_c)/w_2 $.   Substituting this in~\eqref{eq:rinf}
and simplifying  yields
\[
         \tilde   R ( \lambda_c)=\begin{cases}
                           \frac{   (b-w_1 \lambda_c) ( (w_1+w_2)\lambda_c-b  )    }{ w_2^2\lambda_c }, &   \lambda_c > 2b/(2w_1+w_2), \\
                           \lambda_c/4,              &  \lambda_c \leq  2b/(2w_1+w_2).
            \end{cases}
\]
This is a concave function of~$\lambda_c$ and its unique maximum can be obtained by
differentiating with respect to~$\lambda_c$ and equating the derivative to zero.
This yields~$\tilde \lambda_c^*$ in~\eqref{eq:opt_inf}. Using~$ w_1 \lambda_c +w_2 \lambda =b $
yields~$\tilde \lambda^*$, and substituting~$\tilde \lambda^*,\tilde \lambda^*_c$ in~\eqref{eq:rinf} completes the proof.~$\square$

\bibliographystyle{IEEEtranS}
\bibliography{RFM_bibl}

\begin{IEEEbiography}[{\includegraphics[width=1.25in,height=1.25in,clip,keepaspectratio]{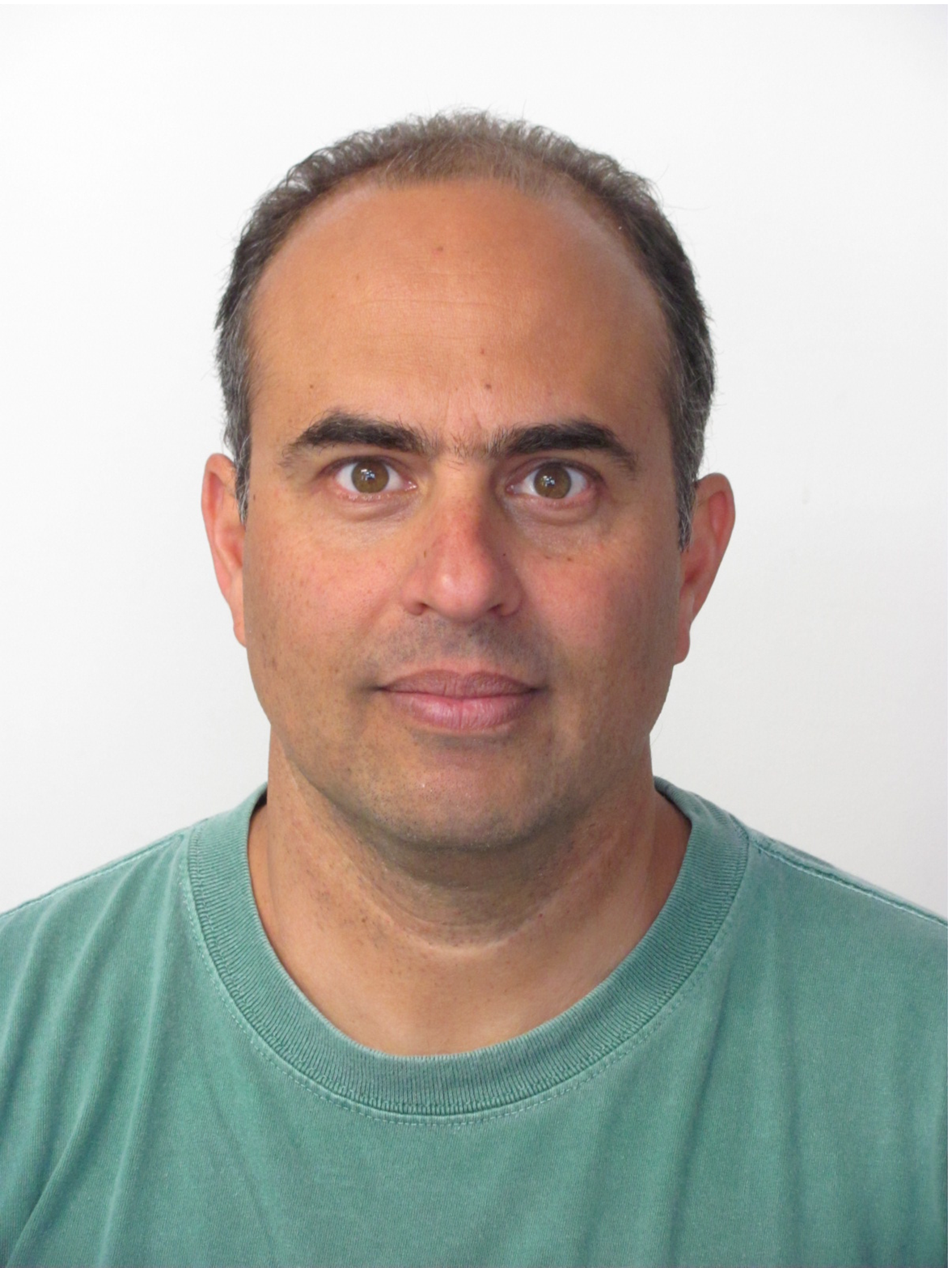}}]{Yoram Zarai}
received the B.Sc. (cum laude) and M.Sc. degrees in Electrical Engineering from Tel Aviv University, in 1992 and 1998 respectively. He is currently working toward the PhD degree at Tel Aviv University. His research interests include modeling and analysis of biological phenomena, machine learning and signal processing.
\end{IEEEbiography}
\begin{IEEEbiography}[{\includegraphics[width=1.25in,height=1.25in,clip,keepaspectratio]{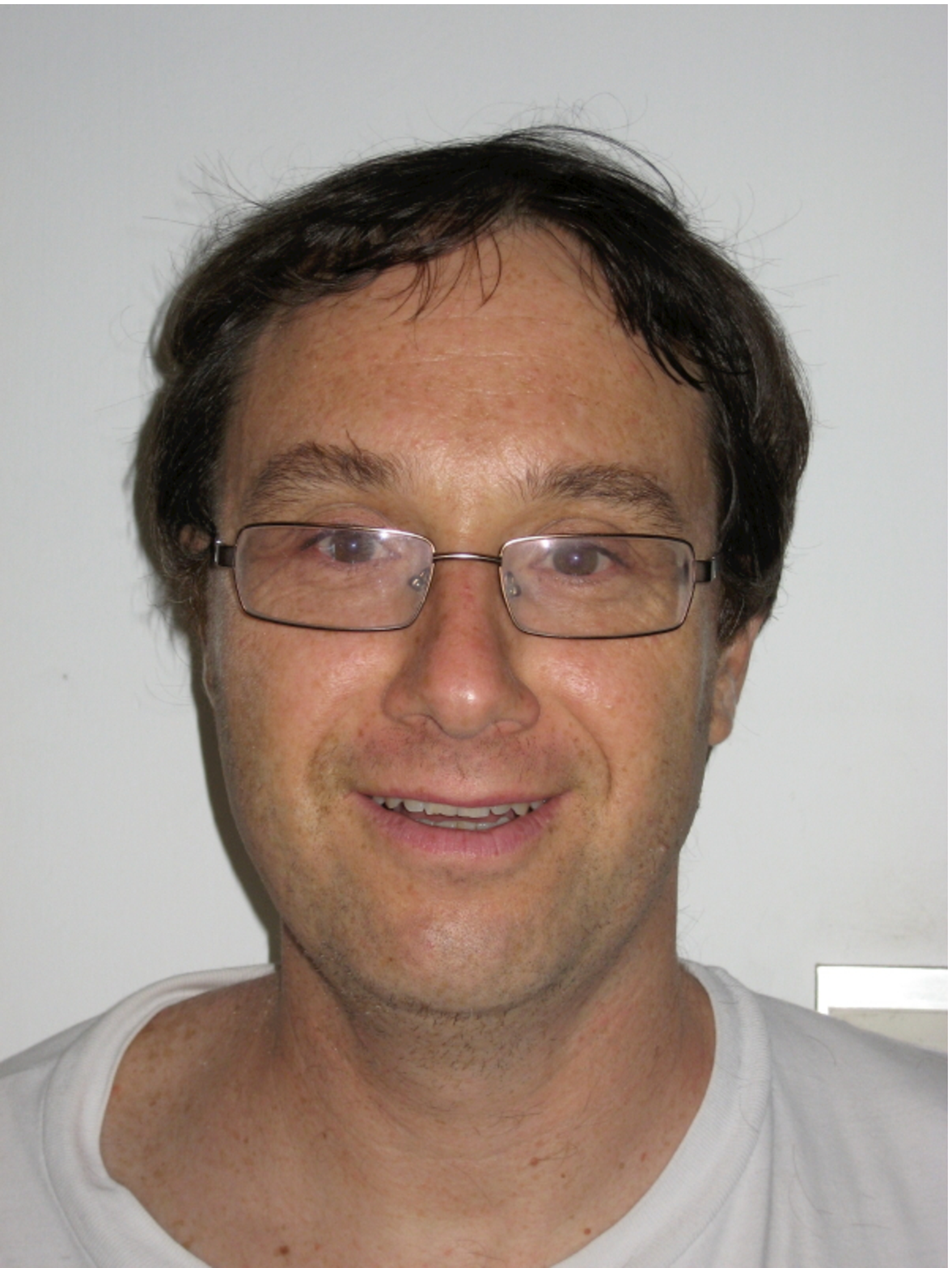}}]{Michael Margaliot}
received the B.Sc. (cum laude) and M.Sc. degrees in Electrical Engineering from the Technion Ð Israel Institute of Technology Ð in 1992 and 1995, respectively, and the Ph.D. degree (summa cum laude) from Tel Aviv University in 1999. He was a post-doctoral fellow in the Department of Theoretical Mathematics at the Weizmann Institute of Science, Rehovot, Israel.  In 2000, he joined the faculty of the Department of Electrical Engineering-Systems, Tel Aviv University, where he is currently an Associate Professor.
His research interests include stability theory, switched systems, optimal control theory, Boolean control networks, fuzzy modeling and control, and systems biology. He is co-author of \emph{New Approaches to Fuzzy Modeling and Control: Design and Analysis}
 (World Scientific, 2000) and of \emph{Knowledge-Based Neurocomputing: A Fuzzy Logic Approach} (Springer, 2009).
\end{IEEEbiography}
\begin{IEEEbiography}[{\includegraphics[width=1.25in,height=1.3in,clip,keepaspectratio]{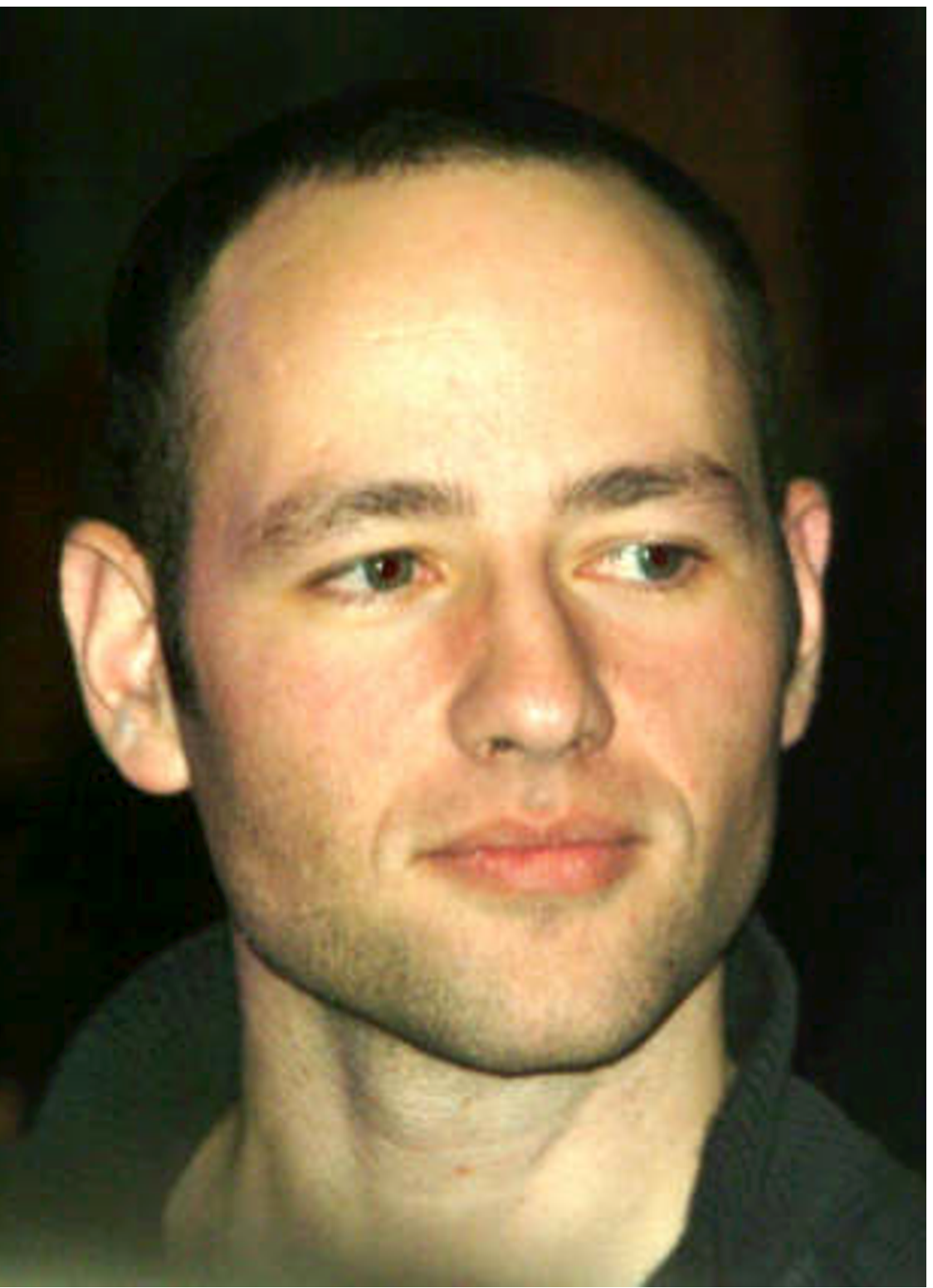}}]{Tamir Tuller}
received the B.Sc. degree in electrical engineering, mechanical engineering and computer science from Tel Aviv University, Tel Aviv, Israel, the M.Sc. degree in electrical engineering from the Technion- Israel Institute of Technology, Haifa, Israel, and Ph.D. degrees in computer science and medical science from Tel Aviv University. He was a Safra Postdoctoral Fellow in the School of Computer Science and the Department of Molecular Microbiology and Biotechnology at Tel Aviv University, and a Koshland Postdoctoral Fellow in the Faculty of Mathematics and Computer Science in the Department of Molecular Genetics at the Weizmann Institute of Science, Rehovot, Israel. In 2011, he joined the Department of Biomedical Engineering, Tel Aviv University, where he is currently an Assistant Professor. His research interests fall in the general areas of computational biology, systems biology, and bioinformatics. In particular, he works on deciphering, computational modeling, and engineering of gene
 expression.\end{IEEEbiography}

 \end{document}